\documentclass[preprint,12pt]{elsarticle}
\usepackage[T1]{fontenc}
\input{glyphtounicode}
\usepackage[utf8]{inputenc}
\usepackage{lmodern}
\usepackage{microtype}
\usepackage{amsmath,amssymb,mathtools}
\usepackage{booktabs,array}
\usepackage{enumitem}
\usepackage{tikz}
\usetikzlibrary{arrows.meta}
\usepackage[hidelinks]{hyperref}
\hypersetup{pdftitle={An Exact Rounds-versus-Queries Trade-off for Pointer Chasing,
  Machine-Checked in Lean 4},pdfauthor={Rafig Huseynzade}}
\usepackage{newunicodechar}
\newunicodechar{∧}{\ensuremath{\wedge}}
\newunicodechar{∀}{\ensuremath{\forall}}
\newunicodechar{∃}{\ensuremath{\exists}}
\newunicodechar{≠}{\ensuremath{\neq}}
\newunicodechar{≤}{\ensuremath{\leq}}

\newtheorem{theorem}{Theorem}[section]
\newtheorem{proposition}[theorem]{Proposition}
\newtheorem{corollary}[theorem]{Corollary}
\newtheorem{lemma}[theorem]{Lemma}
\newdefinition{definition}[theorem]{Definition}
\newdefinition{example}[theorem]{Example}
\newdefinition{remark}[theorem]{Remark}
\newproof{proof}{Proof}
\newproof{proofof}{Proof of}

\newcommand{\Cells}{\mathcal{C}}
\newcommand{\out}{\mathsf{out}}
\newcommand{\ask}{\mathsf{ask}}
\newcommand{\eval}{\mathsf{eval}}
\newcommand{\cost}{\mathrm{cost}}
\newcommand{\lean}[1]{\texttt{#1}}

\tikzset{
  cell/.style={draw, minimum size=5.8mm, inner sep=0pt, font=\small},
  unread/.style={cell, draw=black!35, text=black!35, densely dotted},
  hit/.style={cell, fill=black!12, very thick},
  front/.style={cell, very thick, draw=black, double, double distance=0.8pt, text=black!35},
  link/.style={-{Stealth[length=2.2mm]}, thick},
}

\begin{document}

\begin{frontmatter}

\title{An Exact Rounds-versus-Queries Trade-off for Pointer Chasing,\\ Machine-Checked in Lean~4}

\author{Rafig Huseynzade}
\ead{huseynzaderafig@gmail.com}
\address{Independent Researcher, Baku, Azerbaijan}

\begin{abstract}
A deterministic query algorithm with $d$ rounds submits $d$ batches of queries, each chosen from the
answers to earlier batches. Canonne and Gur asked for general bounds on the cost of removing rounds.
For $k$-step pointer chasing over $k$ tables of $m\ge2$ entries we determine this cost exactly: for
$1\le d\le k$, the least worst-case number of queries with at most $d$ rounds is exactly
$(k-d)m+d$, so every round removed costs exactly $m-1$ queries, which is what the naive merging of
two consecutive rounds pays. The lower bound uses an adversary that answers every query with the
queried index, except at the first unread cell of the chain. The model, the problem and the theorem
are formalised in Lean~4 without Mathlib, with no \texttt{sorry} and only the axioms
\texttt{propext} and \texttt{Quot.sound}; an exhaustive search confirms the formula for small
parameters.
\end{abstract}


\end{frontmatter}

\section{Introduction}\label{sec:intro}

A query algorithm that may choose each query after seeing earlier answers can be far cheaper than
one that must fix its queries in advance, and between the two extremes lies the number of
\emph{rounds} in which the algorithm may adapt. Canonne and Gur~\cite{CanonneGur2018} asked for
general bounds on what removing rounds must cost (their Open Problem~3). This paper gives an exact
answer at every number of rounds for pointer chasing: the cost of removing a round equals the cost
of the naive simulation that merges two consecutive rounds, so the simulation cannot be improved.

Many computations follow a trail. A program walks a linked list stored in remote memory, a query
follows a chain of references through a database, a parallel algorithm traces a path through a
successor table. Following a trail of length $k$ one step at a time needs $k$ round trips, because the
address of each step is known only after the previous one has been read. When round trips are
expensive (network latency, synchronisation between parallel workers, a storage system that serves
requests in batches) one would rather fetch more in each round and use fewer rounds. The question is
the price: \emph{how many additional reads does each saved round cost?}

This paper answers the question exactly for the simplest trail problem, pointer chasing, in the
standard model of deterministic query algorithms that read their input in rounds (decision trees of
bounded adaptivity~\cite{CanonneGur2018}). We give the model a self-contained formulation, suited to
formalisation. A machine never sees its input directly. It reads the input through an interface
$\iota$ that, given a cell $c$, returns its content $x(c)$.
Reads happen in \emph{rounds}: in each round the machine submits a batch of cells, chosen as a
function of the answers of earlier rounds only, and receives all the answers of the batch together.
The \emph{depth} of a machine is the number of rounds it uses, and its \emph{cost} is the number of
cells it reads. Every cell can be read in any round, so depth restricts \emph{how} the machine reads
its input, never \emph{what} it may read.

\paragraph{The problem} In $k$-step pointer chasing the input consists of $k$ tables
$T_0,\dots,T_{k-1}$, each mapping $\{0,\dots,m-1\}$ to itself. Starting from $a_0=0$ the chain is
$a_1=T_0(a_0)$, $a_2=T_1(a_1)$, and so on, and the input is accepted if the last value
$T_{k-1}(a_{k-1})$ is non-zero. With $k$ rounds the problem is easy: read one cell per round,
$k$ reads in total. With fewer rounds an algorithm must read some cells before it knows whether they
lie on the chain.

\paragraph{The result} We determine the cost at every depth (Theorem~\ref{thm:main}):
\[
\cost_d(L_k) \;=\; (k-d)\,m + d \qquad (1\le d\le k),
\]
where $\cost_d(L_k)$ is the least cost with at most $d$ rounds (Definition~\ref{def:depth}),
and no algorithm with $0$ rounds decides the problem. The formula has a direct reading. With $d$
rounds an algorithm can afford to follow $d$ steps of the chain one cell at a time; each of the other
$k-d$ steps must be covered by reading an entire table, because the algorithm does not yet know where
the chain will enter it. Each round removed therefore costs exactly $m-1$ extra reads, and for $k\ge2$
the step from $k$ rounds to $k-1$ rounds raises the cost from $k$ to $m+k-1$. In terms of the input
length $N$ in bits, the cost with $k-1$ rounds is at least $N/(k(\log_2 N+1))$ against $k$ with $k$
rounds (Corollary~\ref{cor:hier}).

The upper bound is the algorithm just described. The lower bound needs an adversary, and the obvious
one fails: an adversary that answers $0$ to every query off the chain lets an algorithm read a whole
table, see that it is constant, and learn where the chain leaves the table without knowing where it
enters. Our adversary answers every query $(j,a)$ with $a$, so that every table it reveals is the
identity, and a completely read table says nothing about the chain. The exception is the first
unread cell of the chain, which the adversary uses, whenever that cell is read, to steer the chain
into a cell the algorithm has not read. The proof fits in two pages (Section~\ref{sec:lower}).

\paragraph{Contributions}
\begin{enumerate}[leftmargin=1.6em,itemsep=2pt]
  \item A self-contained formulation, suited to formalisation, of deterministic query algorithms with
        a bounded number of rounds, in which depth counts rounds (batches of queries) and nothing else
        (Section~\ref{sec:model}).
  \item The exact trade-off between rounds and reads for pointer chasing, at every depth
        (Theorem~\ref{thm:main}), with elementary proofs of both halves
        (Sections~\ref{sec:upper} and~\ref{sec:lower}).
  \item A formalisation in Lean~4 of the model, the problem and the theorem itself, both halves and
        every depth, checked by the Lean kernel using only the axioms \texttt{propext} and
        \texttt{Quot.sound} (Section~\ref{sec:lean}).
  \item An independent computational confirmation: an exhaustive minimax search over all strategies
        for small parameters, and an exhaustive test of the adversary (Appendix~\ref{app:search}).
\end{enumerate}

\paragraph{Scope} That rounds of adaptivity matter is well known. Canonne and Gur proved an
adaptivity hierarchy for property testing, where algorithms are randomised and approximate; on the way
they show that pointer chasing within a single array separates consecutive numbers of rounds for
decision trees by a polynomial gap~\cite{CanonneGur2018}. Pointer chasing is also the standard
example separating $k$-round from $(k-1)$-round protocols in communication
complexity~\cite{NisanWigderson1993,PRV2001}. The setting here is deterministic and exact, and the
proof is elementary. Exact round-versus-query results exist for other search problems
(Section~\ref{sec:related}). What this paper adds is the exact value at every depth for pointer
chasing, the observation that it matches the merging simulation of two rounds (so that, for this
problem, the naive merging bound cannot be improved), and a proof simple enough to be checked by a
machine from the definitions up. The input is accessed only through the interface,
and the measures count rounds and reads, not running time: the result is a statement about query
complexity only.

\paragraph{Organisation} Section~\ref{sec:model} defines the model, Section~\ref{sec:witness} the
problem, and Section~\ref{sec:results} states the results. Sections~\ref{sec:upper}
and~\ref{sec:lower} prove the upper and the lower bound. Section~\ref{sec:lean} describes the
formalisation, Section~\ref{sec:related} discusses related work and Section~\ref{sec:directions}
lists open directions. Appendix~\ref{app:search} reports the computational checks.

\section{The model}\label{sec:model}

Fix a finite set $\Cells$ of \emph{cells} and a finite set $V$ of \emph{values} containing a
designated element $0$. An \emph{input} is a function $x:\Cells\to V$, and a \emph{language} is a
set of inputs. A machine accesses $x$ only through the interface $\iota(c)=x(c)$.

For a finite list $Q$ of cells, the \emph{restriction} of $x$ to $Q$ is the function
\[
x|_Q(c)=\begin{cases} x(c) & \text{if } c\in Q,\\ 0 & \text{otherwise.}\end{cases}
\]
It records the answers to the queries in $Q$ and nothing else.

\begin{definition}[Strategies]\label{def:strat}
Strategies are the trees generated by two constructors:
\begin{itemize}[leftmargin=1.4em,itemsep=1pt]
  \item $\out(b)$, which halts with verdict $b\in\{0,1\}$;
  \item $\ask(Q,f)$, which reads the cells of the list $Q$ in one round and continues with the
        strategy $f(y)$, where $y=x|_Q$ is the restriction of the input to $Q$; here $f$ is any
        function from restrictions to strategies.
\end{itemize}
The \emph{verdict} of a strategy on an input is
$\eval(\out(b),x)=b$ and $\eval(\ask(Q,f),x)=\eval(f(x|_Q),x)$.
A strategy \emph{decides} a language $L$ if $\eval(s,x)=1 \iff x\in L$ for every input $x$.
\end{definition}

The cells of a round are fixed before any of their answers is known, and the rest of the
computation may depend on those answers in any way. Adaptivity is therefore measured exactly by the
number of rounds.

\begin{definition}[Depth and cost]\label{def:depth}
A strategy $s$ has \emph{depth at most $d$} if $s=\out(b)$, or if $s=\ask(Q,f)$, $d\ge1$, and every
continuation $f(y)$ has depth at most $d-1$. It has \emph{cost at most $B$} if $s=\out(b)$, or if
$s=\ask(Q,f)$, $|Q|\le B$, and every continuation $f(y)$ has cost at most $B-|Q|$. For a language $L$
and $d\ge0$, let $\cost_d(L)$ be the least $B$ such that some strategy of depth at most $d$ and cost
at most $B$ decides $L$, and $\cost_d(L)=\infty$ if there is none.
\end{definition}

Depth and cost quantify over every continuation, so they are properties of the strategy tree, in the
way that the depth of a decision tree is. This does not change $\cost_d(L)$: replacing every
continuation that no input reaches by $\out(0)$ changes no verdict, and afterwards the depth and cost
of the strategy are the largest number of rounds and of reads on a single input. Hence $\cost_d(L)$
is also the least worst-case number of reads, over inputs, of a strategy with at most $d$ rounds.
Cost counts the length of each batch, so a cell read twice is paid for twice. The proof of the lower
bound in Section~\ref{sec:lower} counts distinct cells, so it does not depend on this convention.

Two extreme cases are familiar. With $d=1$, a strategy chooses all its queries in advance:
$\cost_1(L)$ is the non-adaptive query complexity of $L$. At the other end, let $D(L)$ be the
deterministic query (decision-tree) complexity of $L$. A decision tree of depth $D(L)$ is a strategy
that reads one cell per round, and reading the cells of a batch one at a time turns any strategy of
cost $B$ into a decision tree of depth at most $B$; hence $\cost_d(L)=D(L)$ for every $d\ge D(L)$. The
quantities $\cost_d(L)$ interpolate between the two.

Our depth counts every batch, including the first, which is chosen before any answer is known. In the
terminology of rounds of adaptivity used by Canonne and Gur~\cite{CanonneGur2018}, where the first
batch is not counted, a strategy of depth $d$ has $d-1$ rounds of adaptivity; thus depth $1$ is
non-adaptive.

\begin{remark}[Machines]\label{rem:tm}
The strategies above describe the query behaviour of a Turing machine equipped with an address tape
and an answer tape. The machine writes a batch of addresses, enters a distinguished
\textsc{round} state, and finds the values of the addressed cells on the answer tape; it has no other
access to the input. Its depth is the number of \textsc{round} transitions on a run, and its cost is the
number of addresses submitted. Since its next batch depends only on the answers it has received, its
behaviour is a strategy in the sense of Definition~\ref{def:strat}. Lower bounds on strategies are
therefore lower bounds on such machines, whatever their running time. The strategies we use for the
upper bounds are easily carried out by such machines in time polynomial in the input length; we do
not analyse running time further, and the formalisation of Section~\ref{sec:lean} is about strategies.
\end{remark}

\section{The problem: pointer chasing}\label{sec:witness}

Fix integers $k\ge1$ and $m\ge2$, and write $[m]=\{0,1,\dots,m-1\}$. The cells are the pairs
$(j,a)$ with $0\le j<k$ and $a\in[m]$, cell $(j,a)$ being column $a$ of table $j$, and the values are
$V=[m]$. An input $x$ is thus a list of $k$ tables $T_j=x(j,\cdot):[m]\to[m]$. Encoded in binary, an
input has $N=k\,m\,\lceil\log_2 m\rceil$ bits.

\begin{definition}[$k$-step pointer chasing]\label{def:lk}
For an input $x$, let $a_0=0$ and $a_{j+1}=x(j,a_j)$ for $0\le j<k-1$. The input belongs to $L_k$
if and only if $x(k-1,a_{k-1})\ne0$. We call $a_0,\dots,a_{k-1}$ the \emph{chain} of $x$ and
$(j,a_j)$ its \emph{chain cells}. The \emph{label} of $x$ is $1$ if $x\in L_k$ and $0$ otherwise; for
a statement $P$ we write $[P]$ for $1$ if $P$ holds and $0$ otherwise, so the label of $x$ is
$[x(k-1,a_{k-1})\ne0]$.
\end{definition}

The chain visits each table exactly once, so no cell is visited twice. Every function
$\Cells\to[m]$ is a valid input. Which cells matter depends on the input: only the $k$ chain cells
determine the label, but where they are is known only by following the chain.

\begin{figure}[t]
\centering
\begin{tikzpicture}[x=8mm,y=-10mm]
  \foreach \a in {0,1,2,3} \node[font=\scriptsize] at (\a,-0.75) {\a};
  \node[anchor=east] at (-0.6,0) {$T_0$};
  \node[anchor=east] at (-0.6,1) {$T_1$};
  \node[anchor=east] at (-0.6,2) {$T_2$};
  \foreach \a/\v in {0/2,1/0,2/3,3/1} \node[cell] (c0\a) at (\a,0) {\v};
  \foreach \a/\v in {0/1,1/3,2/3,3/0} \node[cell] (c1\a) at (\a,1) {\v};
  \foreach \a/\v in {0/0,1/2,2/1,3/2} \node[cell] (c2\a) at (\a,2) {\v};
  \node[hit] at (0,0) {2};
  \node[hit] at (2,1) {3};
  \node[hit] at (3,2) {2};
  \draw[link] (c00.south) -- (c12.north);
  \draw[link] (c12.south) -- (c23.north);
  \node[anchor=west,font=\footnotesize] at (3.7,0) {$a_1=x(0,0)=2$};
  \node[anchor=west,font=\footnotesize] at (3.7,1) {$a_2=x(1,2)=3$};
  \node[anchor=west,font=\footnotesize] at (3.7,2) {$x(2,3)=2\ne0$};
\end{tikzpicture}
\caption{An input of $L_3$ with $m=4$. Row $j$ is the table $T_j$; the numbers above the
columns are the column indices. The chain starts in column $0$ of $T_0$ and each value is the column
to visit in the next table. The shaded chain cells are the only cells that matter for this input.}
\label{fig:instance}
\end{figure}

\begin{example}\label{ex:instance}
Figure~\ref{fig:instance} shows an input with $k=3$ and $m=4$. Its chain is $a_0=0$, $a_1=2$,
$a_2=3$, and its last chain cell holds $2\ne0$, so it belongs to $L_3$. Changing any cell other than
the three shaded ones does not change the label, but changing $x(0,0)$ to $1$ moves the rest of the
chain to column $1$ of $T_1$ and then to column $3$ of $T_2$, where the value $2$ again gives
acceptance; changing $x(0,0)$ to $3$ instead leads to $x(1,3)=0$, then $x(2,0)=0$, and rejection.
\end{example}

\section{Results}\label{sec:results}

\begin{theorem}[Exact trade-off]\label{thm:main}
For all $k\ge1$, $m\ge2$ and $d\ge0$,
\[
\cost_d(L_k)=\begin{cases}
\infty & \text{if } d=0,\\
(k-d)\,m+d & \text{if } 1\le d\le k,\\
k & \text{if } d\ge k.
\end{cases}
\]
\end{theorem}

The two formulas agree at $d=k$. The upper bound is Proposition~\ref{prop:upper} and the lower bound
is Proposition~\ref{prop:lower}.

\begin{corollary}[Depth hierarchy]\label{cor:hier}
Let $k\ge1$ and $m\ge2$.
\begin{enumerate}[label=(\roman*),leftmargin=2em,itemsep=1pt]
  \item $\cost_k(L_k)=k$; $\cost_{k-1}(L_k)=m+k-1$ if $k\ge2$, and $\cost_{k-1}(L_k)=\cost_0(L_1)=\infty$
        if $k=1$.
  \item For $2\le d\le k$, $\cost_{d-1}(L_k)-\cost_d(L_k)=m-1$: every round removed costs exactly
        $m-1$ additional reads.
  \item In terms of the input length $N=km\lceil\log_2m\rceil$, for $k\ge2$,
        $\cost_{k-1}(L_k)\ge m\ge N/\bigl(k(\log_2N+1)\bigr)$, while $\cost_k(L_k)=k$.
\end{enumerate}
\end{corollary}

\begin{proof}
Parts (i) and (ii) are immediate from Theorem~\ref{thm:main}. For (iii), $\lceil\log_2 m\rceil\le
\log_2 m+1\le\log_2N+1$, so $m=N/(k\lceil\log_2m\rceil)\ge N/(k(\log_2N+1))$.
\end{proof}

\begin{remark}\label{rem:where}
The strategy $U_d$ of Section~\ref{sec:upper} reads the last $k-d$ tables completely. For these
tables it does not matter \emph{when} they are read, only \emph{how many} are read: the chain
reaches them only after round $d$, so they may be read in the first round, in the last round, or
spread over several rounds. (Reading the complete tables in round~1 works for any choice of $k-d$
tables other than $T_0$, but a table in the middle of the chain must be read before the chain reaches
it.) What the lower bound shows is that there is no cheaper way to make up for a missing round than
reading a whole table, not even reading parts of several tables.
\end{remark}

\section{The upper bound}\label{sec:upper}

\begin{proposition}[Upper bound]\label{prop:upper}
For every $1\le d\le k$ there is a strategy of depth at most $d$ and cost at most $(k-d)m+d$ that
decides $L_k$. Consequently $\cost_d(L_k)\le(k-d)m+d$ for $1\le d\le k$, and $\cost_d(L_k)\le k$ for
$d\ge k$.
\end{proposition}

\begin{proof}
The strategy $U_d$ works as follows.
\begin{itemize}[leftmargin=1.4em,itemsep=1pt]
  \item In round $r$, for $r=1,\dots,d-1$, read the single cell $(r-1,a_{r-1})$. Its value is
        $a_r$, so the address needed in the next round is known when that round begins.
  \item In round $d$, read the cell $(d-1,a_{d-1})$ together with all $m$ cells of each of the
        tables $d,d+1,\dots,k-1$.
  \item Output the value $[x(k-1,a_{k-1})\ne0]$, computed as follows.
\end{itemize}
If $d=k$, round $d$ reads the last chain cell $(k-1,a_{k-1})$ itself. If $d<k$, after round $d$ the
value $a_d=x(d-1,a_{d-1})$ is known; since tables $d,\dots,k-1$ have been read completely, the
strategy can compute the remaining chain values $a_{d+1},\dots,a_{k-1}$ (none if $d=k-1$) and finally
$x(k-1,a_{k-1})$ from the answers, without further reads. The strategy uses $d$ rounds and reads
$(d-1)+1+(k-d)m$ cells on every input. For $d\ge k$ the strategy $U_k$ has depth at most $d$ and cost
$k$.
\end{proof}

\begin{figure}[t]
\centering
\begin{tikzpicture}[x=8mm,y=-8.5mm]
  \foreach \a in {0,1,2,3} \node[font=\scriptsize] at (\a,-0.7) {\a};
  \foreach \j in {0,1,2,3} {
    \node[anchor=east] at (-0.6,\j) {$T_\j$};
    \foreach \a in {0,1,2,3} \node[unread] (u\j\a) at (\a,\j) {};
  }
  \node[hit] (r00) at (0,0) {1};
  \node[hit] (r12) at (2,1) {2};
  \foreach \j in {2,3} \foreach \a in {0,1,2,3} \node[cell,fill=black!12] at (\a,\j) {2};
  \draw[link] (r00.south) -- (r12.north);
  \node[anchor=west,font=\footnotesize,align=left] at (3.7,0) {round 1: $(0,a_0)$, $a_1=2$};
  \node[anchor=west,font=\footnotesize,align=left] at (3.7,1) {round 2: $(1,a_1)$};
  \node[anchor=west,font=\footnotesize,align=left] at (3.7,2.5) {round 2: all of $T_2$, $T_3$};
\end{tikzpicture}
\caption{The strategy $U_d$ of Proposition~\ref{prop:upper} for $k=4$, $m=4$, $d=2$. The number in a
cell is the round in which it is read; shaded cells are read, dotted cells are never read. Round~1
reads the first chain cell, whose value (here $2$) tells round~2 where to look in $T_1$. Round~2 reads
that cell and the last $k-d=2$ tables in full, after which the rest of the chain can be followed
without reading. Total: $(k-d)m+d=10$ reads.}
\label{fig:upper}
\end{figure}

Figure~\ref{fig:upper} illustrates the strategy. Other choices of the complete tables work equally
well (Remark~\ref{rem:where}); for instance, round~1 may read $(0,0)$ together with any $k-d$ complete
tables other than $T_0$, and rounds $2,\dots,d$ then follow the chain, one cell per round, through the
$d-1$ remaining tables not read in full. Merging the last two rounds of $U_d$, that is, reading in
round $d-1$ every cell that round $d$ might read, gives exactly $U_{d-1}$: the extra cost is the $m-1$
other cells of table $d-1$.

\section{The lower bound}\label{sec:lower}

\begin{proposition}[Lower bound]\label{prop:lower}
Let $d\ge0$ and let $s$ be a strategy of depth at most $d$ that decides $L_k$. Then $d\ge1$, and
there are an input $x$ and an integer $e$ with $1\le e\le\min(d,k)$ such that $s$ reads at least
$(k-e)m+e$ distinct cells on $x$. Consequently the cost of $s$ is at least $(k-d)m+d$ if $d\le k$, and
at least $k$ if $d\ge k$.
\end{proposition}

Together with Proposition~\ref{prop:upper} this proves Theorem~\ref{thm:main}. Before the proof we
explain why the adversary used for weaker round lower bounds does not suffice.

\subsection{Why the obvious adversary fails}\label{sec:obvious}

A standard way to show that $k-1$ rounds are not enough is to answer every query off the chain with
$0$, and to answer the chain cell read in a round with a column $w$ such that cell $w$ of the next
table has not been read, counting the cells of the current round. This is possible as long as that
table has not been read in full. The chain then advances by at most one table per round, so a
strategy with at most $k-1$ rounds that reads fewer than $m$ cells never reads the last chain cell.
This argument shows that $k-1$ rounds need at least $m$ reads, but it cannot give the exact bound.
Take $k=2$ and $d=1$: the exact cost is $m+1$, yet against the zero adversary the $m$ reads of table
$T_1$ suffice, because the strategy sees a constant table and knows that $x(1,a_1)=0$ whatever $a_1$
is. A table read in full and found constant tells the strategy where the chain leaves it without
telling it where the chain enters. The adversary must never reveal a table of this kind.

\subsection{The identity adversary}\label{sec:identity}

The adversary below keeps a complete input $x$, which it modifies during the run, and a
\emph{frontier} $(t,v)$ with $0\le t\le k$. While $t<k$, the frontier is the first chain cell of $x$
that the strategy has not read. Initially $x$ is the identity input, $x(j,a)=a$ for every cell, and
the frontier is $(0,0)$.

In each round, the strategy submits a batch $Q$, chosen from the answers it has received so far. The
adversary then proceeds as follows.
\begin{enumerate}[leftmargin=1.6em,itemsep=2pt]
  \item If $t<k$ and the frontier cell $(t,v)$ belongs to $Q$, let $R$ be the set of all cells read so
        far, the cells of $Q$ included. Let $t'$ be the least index with $t<t'<k$ such that table
        $t'$ is not contained in $R$, and $t'=k$ if there is none. If $t'<k$, let $w$ be the least
        column with $(t',w)\notin R$, set $x(t,v):=w$ and move the frontier to $(t',w)$. If $t'=k$,
        set $x(t,v):=0$ and let the frontier leave the tables ($t:=k$); from then on step~1 never
        applies again.
  \item In every case, answer each cell of $Q$ with its value in the current $x$.
\end{enumerate}
In words: the adversary answers like the identity table everywhere, except that when the strategy
reads the frontier cell, the adversary sends the chain to the first table after $T_t$ that still has
an unread cell, and to the least unread cell of that table. Because every table the chain crosses on
its way there has been read in full and is the identity, the chain keeps the value $w$ while crossing
it.

\begin{lemma}[Invariants]\label{lem:inv}
After every round the following hold.
\begin{enumerate}[label=(\alph*),leftmargin=2em,itemsep=1pt]
  \item Every answer given so far is the value of its cell in the current $x$.
  \item If $t<k$: every cell of every table $j>t$ holds its own column, $x(j,a)=a$; the frontier cell
        holds $x(t,v)=v$; the frontier cell has not been read; and the chain of $x$ has $a_t=v$.
  \item Let $e$ be the number of times the frontier has moved. The tables at which the frontier stood
        before each move, the \emph{stop tables}, are $e$ distinct tables $j<t$, and the chain cell
        $(j,a_j)$ of each of them has been read. Every other table $j<t$ is \emph{crossed}: all $m$ of
        its cells have been read.
  \item The frontier has moved at most once per round.
\end{enumerate}
\end{lemma}

\begin{proof}
Initially $x$ is the identity, no cell has been read, $t=0$, $e=0$ and $a_0=0=v$, so (a)--(d) hold.
Consider a round with batch $Q$.

If the frontier cell is not in $Q$, then $x$, the frontier and $e$ do not change; the new answers are
values of $x$, so (a) holds, and (b)--(d) are unaffected, the frontier cell remaining unread.

Otherwise the adversary changes $x$ at the single cell $(t,v)$, which had not been read before this
round by (b). Earlier answers therefore remain values of $x$, and the answers of this round are taken
from the updated $x$, so (a) holds. All tables $j>t$ are unchanged, so every table $j>t'$ is still the
identity, and $x(t',w)=w$ if $t'<k$; the cell $(t',w)$ is unread by the choice of $w$. For the chain:
$a_{t+1}=x(t,a_t)=x(t,v)=w$, and for $t<j<t'$ the table $j$ is the identity, so
$a_{j+1}=x(j,a_j)=a_j$; hence $a_{t'}=w$, which proves (b). Table $t$ becomes a stop table, and its
chain cell $(t,v)$ was read in this round; the number of stop tables and the number $e$ of moves both
increase by one, and $t<t'$ keeps the stop tables distinct. Each table $j$ with $t<j<t'$ is crossed,
since it is contained in $R$. Tables below $t$ keep their kind and their chain cells, because their
cells do not change. This proves (c). The frontier moved once in this round, which proves (d).
\end{proof}

Let $x^\ast$ be the input held by the adversary when the strategy halts. By Lemma~\ref{lem:inv}(a)
and induction on the rounds, the strategy run on $x^\ast$ submits the same batches and receives the
same answers as against the adversary, so it reaches the same verdict.

\begin{lemma}[An unfinished run errs]\label{lem:unfinished}
If $s$ halts with the frontier at $(t,v)$ for some $t<k$, then $s$ errs on some input.
\end{lemma}

\begin{proof}
For $u\in\{0,1\}$ let $x_u$ be $x^\ast$ with the frontier cell set to $u$, $x_u(t,v)=u$. The two
inputs differ from $x^\ast$ only at a cell that $s$ has not read (Lemma~\ref{lem:inv}(b)), so on each
of them $s$ receives the same answers as on $x^\ast$ and returns the same verdict $b$. Tables below $t$
are unchanged, so the chain of $x_u$ still has $a_t=v$. If $t=k-1$, the label of $x_u$ is
$[x_u(k-1,v)\ne0]=[u\ne0]$. If $t<k-1$, then $a_{t+1}=u$, and since every table above $t$ is the
identity, $a_{k-1}=u$ and the label is $[x_u(k-1,u)\ne0]=[u\ne0]$. In both cases $x_0\notin L_k$
and $x_1\in L_k$ (recall $m\ge2$), so $b$ is wrong for one of them.
\end{proof}

\begin{proof}[Proof of Proposition~\ref{prop:lower}]
Run $s$ against the adversary and let $x=x^\ast$. Since $s$ decides $L_k$, Lemma~\ref{lem:unfinished}
shows that the frontier has reached $t=k$ when $s$ halts. Let $e$ be the number of times the frontier
moved. It moved from $0$ to $k\ge1$, so $e\ge1$; each move increases $t$, so $e\le k$; and by
Lemma~\ref{lem:inv}(d) it moves at most once per round, while $s$ uses at most $d$ rounds, so $e\le d$.
In particular $d\ge1$.

By Lemma~\ref{lem:inv}(c) with $t=k$, exactly $e$ of the $k$ tables are stop tables, each containing at
least one read cell, its chain cell, and the other $k-e$ tables are crossed, each containing $m$ read
cells. Distinct tables have disjoint cells, so $s$ reads at least $e+(k-e)m$ distinct cells on $x$.

Finally, let $d'=\min(d,k)\ge e$. Then
$\bigl(e+(k-e)m\bigr)-\bigl(d'+(k-d')m\bigr)=(d'-e)(m-1)\ge0$, so $s$ reads at least $(k-d')m+d'$
cells on $x$, which is $(k-d)m+d$ if $d\le k$ and $k$ if $d\ge k$.
\end{proof}

\begin{figure}[t]
\centering
\begin{tikzpicture}[x=8mm,y=-8.5mm]
  \foreach \a in {0,1,2,3} \node[font=\scriptsize] at (\a,-0.7) {\a};
  \foreach \j in {0,1,2,3} \node[anchor=east] at (-0.6,\j) {$T_\j$};
  \node[hit] (f0) at (0,0) {2};
  \foreach \a in {1,2,3} \node[unread] at (\a,0) {\a};
  \foreach \a in {0,1,2,3} \node[cell] (t1\a) at (\a,1) {\a};
  \node[hit] (t12) at (2,1) {2};
  \node[cell] at (0,2) {0};
  \node[cell] at (1,2) {1};
  \node[front] (t22) at (2,2) {2};
  \node[unread] at (3,2) {3};
  \foreach \a in {0,1,2,3} \node[unread] at (\a,3) {\a};
  \draw[link] (f0.south) -- (t12.north);
  \draw[link] (t12.south) -- (t22.north);
  \node[anchor=west,font=\footnotesize] at (3.7,0) {stop table: $x(0,0):=2$};
  \node[anchor=west,font=\footnotesize] at (3.7,1) {crossed: identity};
  \node[anchor=west,font=\footnotesize] at (3.7,2) {new frontier $(2,2)$};
  \node[anchor=west,font=\footnotesize] at (3.7,3) {untouched};
\end{tikzpicture}
\caption{The identity adversary after one round, for $k=4$ and $m=4$, against the batch consisting
of the frontier cell $(0,0)$, all of $T_1$, and the cells $(2,0)$ and $(2,1)$. Solid cells have been
read and show the answers; shaded cells are the chain cells read so far; dotted cells are unread and
show their current value; the double-bordered cell is the new frontier $(2,2)$, which is unread and
holds $2$. Table $T_1$ is read in full, so the adversary sends the chain to $T_2$, the first table
after $T_0$ with an unread cell, by answering $x(0,0)=2$, the least unread column of $T_2$. Through the
identity table $T_1$ the chain keeps its value, $a_2=x(1,2)=2$, and arrives at the new frontier. The
round paid $1+m+2=7$ reads. The $1+m$ reads in $T_0$ and $T_1$ are what the proof of
Proposition~\ref{prop:lower} charges to the stop table $T_0$ and the crossed table $T_1$; the two reads
in $T_2$ only made the adversary place the new frontier in column~$2$ instead of column~$0$. The
strategy has learnt nothing about $a_3=x(2,2)$.}
\label{fig:adversary}
\end{figure}

Figure~\ref{fig:adversary} shows one round of the adversary. The proof shows more than a lower bound
on the worst case: against the identity adversary, \emph{every} strategy that decides $L_k$ pays one
read for each round in which it advances along the chain and $m$ reads for every table the chain
crosses.

\section{Formalisation}\label{sec:lean}

The model and the problem (Definitions~\ref{def:strat}, \ref{def:depth} and~\ref{def:lk}),
Theorem~\ref{thm:main} in all three cases, Proposition~\ref{prop:upper}, and
Proposition~\ref{prop:lower} in the cost form described below, with Lemma~\ref{lem:unfinished} and
Lemma~\ref{lem:inv} (part~(c) in a counting form), have been formalised in Lean~4 (version 4.22.0), using only the core
library, without Mathlib. The development contains no \texttt{sorry} (Lean's placeholder for a
missing proof) and declares no axioms, and the command \texttt{\#print axioms} reports that the main
theorems depend only on \texttt{propext} and \texttt{Quot.sound}, two of the standard axioms of Lean's
logic; in particular they do not use the axiom of choice. The table below lists what is and is not
formalised.

\paragraph{Definitions} Strategies are an inductive type with the two constructors of
Definition~\ref{def:strat}:
{\small
\begin{verbatim}
inductive Strat (C V : Type) where
  | out : Bool → Strat C V
  | ask : List C → ((C → V) → Strat C V) → Strat C V
\end{verbatim}}
The restriction $x|_Q$ is \lean{restrict zero Q x}, the verdict is \lean{eval}, and depth and cost
are inductive predicates \lean{Depth s d} and \lean{Cost s c} following Definition~\ref{def:depth},
with the continuation quantified over all functions $\Cells\to V$ (see the modelling choices
below). Pointer chasing is stated with $n=k-1$ and $M=m$. Cells are \lean{Cell n M}, an abbreviation
for \lean{Fin (n+1) × Fin M}; inputs are functions from cells to \lean{Fin M}; \lean{zero} is the value
$0$ of \lean{Fin M}; and \lean{L x} follows the chain from $a_0=0$ and tests the last value. The
theorems below take $n$ and $M$ as implicit arguments, with an instance \lean{NeZero M}, which follows
from the hypothesis \lean{1 < M} and is needed to name \lean{zero}.

\paragraph{The main theorem} For $1\le d\le k$ the Lean statement reads:
{\footnotesize
\begin{verbatim}
theorem tradeoff (h1 : 1 < M) (d : Nat)
    (hd1 : 1 ≤ d) (hdk : d ≤ n + 1) :
    (∃ s : Strat (Cell n M) (Fin M), Depth s d ∧
        Cost s ((n + 1 - d) * M + d) ∧
        ∀ x, eval zero s x = L x) ∧
    (∀ (s : Strat (Cell n M) (Fin M)) (B : Nat),
        Depth s d → Cost s B →
        B < (n + 1 - d) * M + d →
        ∃ x, eval zero s x ≠ L x)
\end{verbatim}}
The first conjunct is Proposition~\ref{prop:upper} and the second is the cost bound of
Proposition~\ref{prop:lower}. Two further theorems complete Theorem~\ref{thm:main}:
\lean{tradeoff\_deep} states that for $d\ge k$ the least cost is exactly $k$, and \lean{no\_rounds}
that no strategy of depth $0$ decides $L_k$.

\paragraph{The proofs} The upper bound defines the strategy $U_d$ of Section~\ref{sec:upper}
(\lean{ubD}) and proves its depth, its cost and its correctness on every input; correctness rests on
the fact that the answers of the last round agree with the input on every cell the remaining chain
can visit. The lower bound formalises the identity adversary as a recursive function on strategies
(\lean{play}), whose state holds the current input, the frontier, the list of cells read and the
number of moves. Invariant~(b) of Lemma~\ref{lem:inv}, together with a numeric invariant that takes
the place of~(c), forms a structure (\lean{Inv}) preserved by every round (\lean{step\_inv}). The
numeric invariant states that the tables behind the frontier account for at least $e+(t-e)m$ reads
when the frontier has moved $e$ times and stands at table $t$; a crossed table contributes $m$ by a
pigeonhole lemma on lists proved in the development. Invariant~(a) is proved separately, for one round
(\lean{step\_x}) and along the whole run (\lean{play\_agree}), and invariant~(d) is \lean{step\_e} for
one round and \lean{play\_depth} along the run. Lemma~\ref{lem:unfinished} is
\lean{wrong\_of\_frontier}. The Lean form of Proposition~\ref{prop:lower} is \lean{lower\_moves}: if
$s$ has depth at most $d$ and cost at most $B$ and decides $L_k$, then $e+(k-e)m\le B$ for some $e$
with $1\le e\le\min(d,k)$, with no restriction on $d$. It bounds the cost of
Definition~\ref{def:depth}, which counts repeated reads, and this is the form Theorem~\ref{thm:main}
needs. The statement of Proposition~\ref{prop:lower} about distinct cells read on a single input is
proved on paper only. The development also formalises the weaker bound of Section~\ref{sec:obvious},
that with $k-1$ rounds fewer than $m$ reads never suffice, as \lean{lower\_bound}, using a variant of
the zero adversary that fixes the next link after every round.

\paragraph{What the formalisation covers}

\begin{center}\footnotesize
\begin{tabular}{@{}>{\raggedright\arraybackslash}p{0.3\linewidth}>{\raggedright\arraybackslash}p{0.38\linewidth}>{\raggedright\arraybackslash}p{0.22\linewidth}@{}}
\toprule
paper & Lean & status\\
\midrule
Definitions~\ref{def:strat}, \ref{def:depth} & \lean{Strat}, \lean{restrict}, \lean{eval}, \lean{Depth}, \lean{Cost} & definitions\\
Definition~\ref{def:lk} & \lean{follow}, \lean{L} & definition\\
Proposition~\ref{prop:upper} & \lean{ubD\_spec} & proved\\
Proposition~\ref{prop:lower}, cost bound & \lean{lower\_moves}, \lean{exact\_lower}, \lean{no\_rounds} & proved\\
Proposition~\ref{prop:lower}, distinct cells on one input & --- & on paper only\\
Lemma~\ref{lem:inv} & \lean{Inv}, \lean{step\_inv}, \lean{play\_agree}, \lean{play\_depth} & proved, (c) in counting form\\
Lemma~\ref{lem:unfinished} & \lean{wrong\_of\_frontier} & proved\\
Theorem~\ref{thm:main}, $d=0$ & \lean{no\_rounds} & proved\\
Theorem~\ref{thm:main}, $1\le d\le k$ & \lean{tradeoff} & proved\\
Theorem~\ref{thm:main}, $d\ge k$ & \lean{tradeoff\_deep} & proved\\
Corollary~\ref{cor:hier}(i),(ii) & immediate from \lean{tradeoff}, \lean{no\_rounds} & not stated separately\\
Corollary~\ref{cor:hier}(iii) & --- & on paper only (arithmetic)\\
Section~\ref{sec:obvious}, $k-1$ rounds need $m$ reads & \lean{lower\_bound} & proved\\
Section~\ref{sec:model}, $\cost_1$, $\cost_d=D$, pruning & --- & on paper only\\
Remark~\ref{rem:tm} (machines) & --- & not formalised\\
\bottomrule
\end{tabular}
\end{center}

\paragraph{Modelling choices} Two choices in the formal definitions deserve mention, since a
reader checking the statement should know them. First, as in Definition~\ref{def:depth}, depth and
cost quantify over every continuation, including continuations that no input reaches; in Lean a
continuation is moreover defined on all functions $\Cells\to V$, not only on restrictions, and
\lean{Depth} and \lean{Cost} constrain it at every argument. This is the decision-tree convention. It
makes the formal upper bound stronger and the formal lower bound weaker, since the hypotheses
\lean{Depth s d} and \lean{Cost s B} of \lean{exact\_lower} also constrain unreachable branches. The
least cost is the same either way, by the pruning argument of Section~\ref{sec:model}; the proof in
Section~\ref{sec:lower} bounds the reads on the single input $x^\ast$ directly, and the Lean proof uses
the two hypotheses only along the adversary's run (\lean{play\_depth}, \lean{play\_cost}). The pruning
step itself is not formalised. Second, a continuation receives the whole function $x|_Q$, which is $0$
outside $Q$; since $Q$ is part of the strategy, the continuation learns the values of exactly the
cells it asked for.

\section{Related work}\label{sec:related}

\paragraph{Rounds and queries} Trade-offs between rounds and queries have long been studied in
other query models, beginning with Valiant's parallel comparison model~\cite{Valiant1975}. Canonne and
Gur~\cite{CanonneGur2018} proved an adaptivity hierarchy for property testing: for every $k$ up to a
polynomial in the input length there is a property testable with $k$ rounds of adaptivity and
$\tilde O(k)$ queries that requires polynomially more queries with one round fewer. On the way they
prove a hierarchy for decision trees of bounded adaptivity, using pointer chasing within a single
array of length $\ell$ (their ``$k$-iterated address function''): $k+1$ queries suffice with $k$
rounds of adaptivity, while for $k$ up to about $(\ell/\log\ell)^{1/3}$ one round fewer requires
$\Omega(\ell/(k^2\log\ell))$ queries, even for randomised linear decision trees. The lower bound is a
reduction from the communication lower bound of Nisan and Wigderson~\cite{NisanWigderson1993}. Their
rounds of adaptivity do not count the first batch (Section~\ref{sec:model}). Canonne and Gur also bound
the cost of removing a single round by merging two consecutive rounds, querying together with the
first every batch that the second might query, and they ask for general round-reduction bounds for
testers of properties of Boolean functions~\cite[Section~6.1 and Open Problem~3]{CanonneGur2018}. For
pointer chasing, merging the last two rounds of the strategy $U_d$ gives exactly $U_{d-1}$
(Section~\ref{sec:upper}), and Theorem~\ref{thm:main} shows that no strategy does better, at any depth.

Carolan, Gilani and Vempati~\cite{CGV2025} study parallel query complexity, with $p$ queries per round.
The full version of their paper~\cite[Theorem~16 and Appendix~C]{CGV2024full} includes, with a proof
given for completeness, the fact that a deterministic algorithm needs $\Omega(\min(k,b/p))$ rounds,
and $\tilde O(\min(k,b/p))$ suffice, to chase a pointer $k$ steps through a function on $b$ blocks
(the logarithmic factor comes from reading a block bit by bit; with one query per block their
Lemmas~74 and~75 are tight up to a constant). Their adversary also keeps the algorithm from getting
ahead of the chain, but it answers $0$ off the chain; this suffices up to constant factors while few
cells have been read, and it is the adversary that Section~\ref{sec:obvious} shows to be too weak for
an exact count. They bound the number of rounds for a given width, whereas we determine the total
number of queries for a given number of rounds exactly. Girish, Sinha, Tal and Wu~\cite{GSTW2024} prove a strong round hierarchy for
quantum query algorithms, and Li, Fu, Shi and Yao~\cite{LFSY2026} recently determined the two-round
cost of evaluating semantic regular expressions sharply and gave a round hierarchy for it that is
tight up to constant factors. To our knowledge the exact value of $\cost_d(L_k)$ at every depth has not
been stated before. The contribution of this paper is that exact value, an adversary that attains it,
and a formal proof, in a self-contained formulation of the standard round model.

\paragraph{Exact round results for other search problems} Exact values of the price of a round are
known for two search problems. Snir~\cite{Snir1985} showed that a parallel search of a sorted table
of size $n$ with $p$ comparisons per step needs exactly $\lceil\log(n+1)/\log(p+1)\rceil$ steps,
attained by $(p+1)$-ary search. Gerbner and Keszegh~\cite{GerbnerKeszegh2012} showed that finding the
path to a marked leaf of a complete $d$-ary tree of depth $n$ (on $n+1$ levels), with $k$ queries
per round, needs exactly $\lceil n/\log'_d k\rceil$ rounds, where $\log'_d k$ is the largest $i$ with
$1+d+\dots+d^{i-1}\le k$; the obvious algorithm that asks complete levels is optimal, and their
adversary advances the path by at most $\log'_d k$ levels per round. Both results fix the width of a
round and count rounds; in both the cost of a round grows exponentially with the number of levels it
covers, as it does for the tree-pointer function obtained from $L_k$ by giving every chain cell its
own successor table. For $L_k$ itself the cost is linear in the number of rounds removed, because the
tables are shared along the chain: this is what Theorem~\ref{thm:main} determines.

\paragraph{Pointer chasing} Pointer chasing was introduced by Papadimitriou and
Sipser~\cite{PapadimitriouSipser1984} and is the classical witness for round hierarchies in
communication complexity, which go back to Duris, Galil and Schnitger~\cite{DGS1987}. Nisan and
Wigderson~\cite{NisanWigderson1993} showed that $k$-round protocols for $k$-step pointer chasing are
exponentially more efficient than $(k-1)$-round ones, and Ponzio, Radhakrishnan and
Venkatesh~\cite{PRV2001} studied its complexity for every fixed number of rounds in detail. In the
query setting studied here a single party holds the whole input and pays for each cell it reads; the
problem is easier to analyse, and the cost can be pinned down exactly.

\paragraph{Formalised lower bounds} Machine-checked lower bounds in complexity theory have multiplied
recently, mostly for circuits. Thiemann formalised in Isabelle/HOL a superpolynomial lower bound on the
size of monotone circuits computing the clique function~\cite{Thiemann2022}, and Wesonga formalised
H{\aa}stad's lower bound for PARITY in Lean~4~\cite{Wesonga2026}. The Lean~4 library
Complexitylib~\cite{Complexitylib}, built on Mathlib, formalises among much else Schnorr's lower bound
for parity, H{\aa}stad's switching lemma and Nechiporuk's formula bound for rectangle-free families.
For query-type models, Eberl formalised in Isabelle/HOL the $\log_2(n!)$ lower bound on the number of
comparisons of comparison-based sorting~\cite{Eberl2017}, and a Lean library for quantum query
complexity formalises adversary and polynomial-method bounds~\cite{LeeQQC}. We are not aware of a
formalised lower bound for query algorithms with a bounded number of rounds.

\section{Directions}\label{sec:directions}

The model was designed to be extended, and several questions are open.
\begin{itemize}[leftmargin=1.4em,itemsep=2pt]
  \item \emph{The price of a round in general.} If a round reads $q$ cells with values in $V$, and
        the part of the next round that depends on its answers reads $q'$ cells, then merging the two
        rounds costs at most $(|V|^q-1)\,q'$ extra reads. In the strategy $U_d$ we have $q=q'=1$, and
        Theorem~\ref{thm:main} shows that the resulting $m-1$ extra reads per round cannot be
        avoided. Is the bound $(|V|^q-1)\,q'$ attained for larger $q$ and $q'$? Lower bounds of this
        kind, for Boolean values ($|V|=2$), would limit how good a general round-reduction bound can
        be; Canonne and Gur~\cite[Open Problem~3]{CanonneGur2018} ask for such a bound for testers of
        properties of Boolean functions. More broadly: for which languages $L$ is
        $\cost_d(L)$ affine in $d$ on the range $1\le d\le D(L)$, as here, and is there a combinatorial
        measure of a language that determines its whole trade-off curve?
  \item \emph{Pointer chasing in a single array.} Canonne and Gur's $k$-iterated address function
        chases a pointer $k$ times within one array of length $\ell$~\cite{CanonneGur2018}. Its exact
        cost with $d$ rounds is open; a cell may be visited more than once, so the argument of
        Section~\ref{sec:lower} does not apply as it stands.
  \item \emph{Randomised and approximate strategies.} What is the bounded-error cost of $L_k$ with
        $d$ rounds? Is there an equally exact answer, and can it be machine-checked?
  \item \emph{Time.} When the machines of Remark~\ref{rem:tm} are charged for running time as well
        as for reads, do the classes of languages decided in polynomial time with depth $d$ and few reads,
        form a hierarchy in $d$, and with which witnesses?
  \item \emph{Other models of access.} The same questions arise for quantum queries and for
        interfaces that answer richer questions than the content of a cell.
\end{itemize}

\section*{Code availability}

The Lean development of Section~\ref{sec:lean}, four Lean modules under \texttt{lean/} with
\texttt{lakefile.lean} and the pinned \texttt{lean-toolchain}, and the Python script of
Appendix~\ref{app:search}, \texttt{python/pointer\_chasing.py}, are distributed as ancillary files of
the arXiv version of this article (arXiv:2510.08577), together with a \texttt{README.md}; these are
the reference version of the code. The Lean files build with the pinned toolchain (Lean 4.22.0, core
library only) by running \texttt{lake build} in \texttt{lean/}; the script needs Python~3.8 or later
and no third-party packages.

\section*{Use of AI in the research}

An AI system (Claude, Anthropic) was used throughout the research under the direction of the author:
in the design of the model, in finding and writing the proofs, in writing the Lean formalisation and
in writing and running the computational checks. The author verified every argument, and the results
rest on checks that do not depend on the AI's judgement: Theorem~\ref{thm:main},
Propositions~\ref{prop:upper} and~\ref{prop:lower} (the latter in the cost form stated in
Section~\ref{sec:lean}), Lemma~\ref{lem:inv} (part~(c) in the counting form described there) and
Lemma~\ref{lem:unfinished} are checked by the Lean kernel, the table in Section~\ref{sec:lean} lists
the statements proved only on paper, and every computational claim is reproducible with the ancillary
files. The author takes full responsibility for the content.

\section*{Declaration of generative AI and AI-assisted technologies in the writing process}

During the preparation of this work the author used Claude (Anthropic) in order to draft and edit the
text. After using this tool, the author reviewed and edited the content as needed and takes full
responsibility for the content of the publication.

\appendix

\section{Computational confirmation}\label{app:search}

The ancillary script \texttt{pointer\_chasing.py} implements an executable model of strategies and of
$L_k$, used for two independent checks of Theorem~\ref{thm:main}.

\paragraph{Exhaustive minimax over all strategies} For a partial assignment $\alpha$ (the values of
the cells read so far), the set of labels of the inputs that extend $\alpha$ is computed exactly by
propagating the set of reachable addresses from table to table: an address $a$ at table $j$ leads to
$\alpha(j,a)$ if $(j,a)$ has been read and to every address otherwise. This is exact because a chain visits
each table once, so the unread cells along any chain can be chosen independently. A state is
\emph{won} with $r$ rounds and budget $B$ if its label is determined, or if some batch $Q$ with
$|Q|\le B$ leads, for every assignment of values to $Q$, to a state won with $r-1$ rounds and budget
$B-|Q|$; the search may ignore cells that lie on no chain consistent with $\alpha$, since their values
cannot affect the label. This quantifies over all strategies, not over a sample. The least winning
budget equals $(k-d)m+d$ in all $21$ cases with
$(k,m)\in\{(2,2),(2,3),(3,2),(3,3),(4,2),(3,4),(4,3)\}$ and $1\le d\le k$, and no budget wins with
$d=0$ for these seven pairs.

\paragraph{Exhaustive test of the adversary} The identity adversary is deterministic, so against it
every strategy produces a single sequence of batches, and every sequence of batches is produced by
some strategy. Re-reading a cell never moves the frontier and only adds to the cost, so it suffices to
consider batches of distinct, previously unread cells. For all $(k,m,d)$ with $0\le d\le k$ and
\begin{gather*}
(k,m)\in\{(1,2),(1,3),(2,2),(2,3),(2,4),(2,5),(2,6),\\
(3,2),(3,3),(3,4),(4,2),(4,3),(5,2),(6,2)\},
\end{gather*}
every sequence of $d$ such batches of total size below $(k-d)m+d$ leaves the label undetermined, as
Proposition~\ref{prop:lower} predicts; the largest case, $(k,m,d)=(6,2,4)$, has $17{,}713{,}745$ such
sequences. As a control, the same test, run against the zero adversary of Section~\ref{sec:obvious}
(\texttt{zero\_adversary} in the script), finds a sequence below the bound that determines the label for
each $(k,m,d)\in\{(2,2,1),(2,3,1),(3,2,1),(3,3,2)\}$; and for all $1\le k\le5$, $2\le m\le5$ and
$1\le d\le k$, the strategy $U_d$ ends against the identity adversary with the label determined after
exactly $(k-d)m+d$ reads.

\bibliographystyle{elsarticle-num}
\bibliography{references}

\end{document}